\documentstyle[amsfonts,epsfig]{aipproc2}

\newcommand{\ep}{\epsilon}

\newcommand{\La}{\Lambda}

\newcommand{\Om}{\Omega}

\newcommand{\si}{\sigma}

\newcommand{\DD}{\mbox{$\cal D$}}

\newcommand{\SS}{\mbox{$\cal S$}}

\newcommand{\Eq}[1]{ Eq.~(\ref{#1})}
\renewcommand{\baselinestretch}{1.01}
\newcommand{\be}{\begin{equation}}
\newcommand{\ee}{\end{equation}}

\newcommand{\bea}{\begin{eqnarray}}
\newcommand{\eea}{\end{eqnarray}}
\newcommand{\bean}{\begin{eqnarray*}}
\newcommand{\eean}{\end{eqnarray*}}

\newcommand{\bk}{{\bf k}}

\newcommand{\ie}{{\em i.e.~}}
\def\spose#1{\hbox to 0pt{#1\hss}} 
\def\ltapprox{\mathrel{\spose{\lower 3pt\hbox{$\mathchar"218$}} 
 \raise 2.0pt\hbox{$\mathchar"13C$}}} 
\def\gtapprox{\mathrel{\spose{\lower 3pt\hbox{$\mathchar"218$}} 
 \raise 2.0pt\hbox{$\mathchar"13E$}}} 
\def\inapprox{\mathrel{\spose{\lower 3pt\hbox{$\mathchar"218$}} 
 \raise 2.0pt\hbox{$\mathchar"232$}}}

\begin{document}
\title{CMB Anisotropies in Pre-Big Bang Cosmology}

\author{F. Vernizzi, A. Melchiorri and R. Durrer}
\address{D\'epartement de Physique Th\'eorique, 
Universit\'e de Gen\`eve,
24 quai Ernest Ansermet, CH-1211 Gen\`eve 4, Switzerland}

%\lefthead{LEFT head}
%\rig436thead{RIGHT head}
\maketitle

\begin{abstract}
We present an alternative scenario for cosmic structure formation where
initial fluctuations are due to Kalb-Ramond axions produced during a 
pre-big bang phase of inflation. We investigate whether this scenario,
where the fluctuations are induced by seeds and therefore are of
isocurvature nature, can be brought in agreement with present
observations by a suitable choice of cosmological parameters. 
We also discuss several observational
signatures which can distinguish axion seeds from standard inflationary
models. 
\end{abstract}

\section*{Introduction}
The pre-big bang idea (see \cite{PBB} and references therein) represents 
one of the first and most interesting
attempts to develop a new cosmological scenario which solves 
the horizon and flatness problems, based on string theory.
In this radically new picture, the underlying duality 
symmetry present in the low energy sector of string theory
naturally leads 
to an inflationary phase prior
to the big bang during which curvature and dilaton are growing.

Superstring theory is a tensor-scalar theory of gravity 
in 10 dimensions.
The minimal low energy effective 
action of the NS-NS sector of string theory 
can be written under certain conditions as
\begin{equation}
S = \int d^{4}x \sqrt{|g|} e^{-\phi} \left[\mbox{R} +
 (\nabla \phi)^2 -3(\nabla \log b)^2 
-\frac{1}{2}e^{2\phi}(\nabla \sigma)^2 \right],
\label{Action4}
\end{equation}
where we have assumed dimensional reduction by compactification 
of 6 dimensions on some isotropic Ricci flat manifolds. 
This action includes a time dependent dilaton field 
$\phi(t)$. R is the Ricci
scalar of the 4-dimensional metric 
$g_{\mu \nu}=\rm diag[-1,a^2(t),a^2(t),a^2(t)]$ and $\log b(t)$ 
is a  time dependent moduli field, 
where $b(t)$ represents the scale factor of the 6 extra dimensions. 
The last term of this action 
contains a pseudo-scalar field, the axion $\sigma$ 
(not to be confused with the Peccei-Quinn axion), 
which is universal in string theory.

When the axion field is trivial, $\dot{\sigma}=0$, or its 
contribution to the global dynamics of the universe is negligible, 
the cosmological equations derived from (\ref{Action4}), where all the fields 
are supposed to be homogeneous,
are invariant under the
duality transformations,
$a(t) \rightarrow 1/a(-t)$, $\phi(t) \rightarrow 
\phi(-t) - 6\ln(a(-t))$,
which is called {\em scale factor duality} and represents an important
motivation behind the pre-big bang scenario~\cite{PBB}. 
In fact, this property relates
a positive time/post-big bang universe with constant dilaton (our universe)
to a negative time/pre-big bang universe, where 
inflation is driven by the kinetic energy of the growing dilaton. 

For negative time, the field equations for $a,\phi$ and $\log b$ are 
then solved~\cite{PBB}
by the following power laws, known as {\em dilaton-vacuum} solutions: 
\be
a(t)= (-t)^{\alpha}, \ \ \ \ 
b(t) =  
 (-t)^{\beta}, \ \ \ \ 
e^{\phi(t)} = 
	(-t)^{3 \alpha -1},
\label{Solu}
\ee
where $\alpha$ and $\beta$ satisfy  the Kasner constraint,
$3 \alpha^2 + 6 \beta^2 =1$.
From these solutions one can see that, during the pre-big bang phase, \ie
for negative times, 
a negative $\alpha$ and a positive $\beta$ are required 
to make the external 3-dimensional 
space expand and the internal 6-dimensional
space contract. 

\section*{Cosmic structure from axion seeds}

The pre-big bang scenario was thought for some time to be unable 
to provide a scale-invariant spectrum of 
perturbations. First-order tensor and scalar 
perturbations in the metric, together with perturbations in the 
moduli fields and the dilaton,
were found to be characterized by extremely ``blue'' spectra \cite{5b}. 
This large tilt together with a natural normalization 
imposed by the string cutoff at the shortest amplified scales, 
make their contribution to large-scale structure completely negligible.

However, the spectral tilt of the axion
field, if
produced by amplification of vacuum quantum fluctuation 
during the pre-big bang phase,
can assume a whole range of values depending on the behavior 
of the internal and external dimensions; in particular, it 
can naturally provide a scale-invariant spectrum of perturbations 
\cite{Copeland}.

One can compute the primordial axion spectral index 
by varying the action (\ref{Action4}) with respect to the field $\sigma$
and introducing the {\em canonical variable}, 
$\psi\equiv a_A \sigma \equiv ae^{\phi/2}\sigma$, which 
yields the {\em evolution equation} for the axion field,
written in Fourier modes as
\be
 \ddot{\psi_{\bk}} +
(k^2-\ddot{a}_A/ a_{A})\psi_{\bk} =0 . 
\label{Evol}
\ee 
The dot represents the derivative with respects to
conformal time $\eta$, defined by $d\eta=a(t)dt$.

By matching the solution to this equation 
during the pre-big bang phase, properly normalized
to vacuum fluctuation at early times, with the solution found  after the 
singularity in the radiation dominated era, one finds 
the spectrum of the axion field, 
\be
\Omega_{\si}(k)\equiv\frac{1}{\rho_c} \frac{d\rho_{\sigma}(k)}{d\log k} 
\sim g_1^2 \left( \frac{k}{k_1} \right)^{n_{\sigma}-1},
\label{Spettro}
\ee 
where $k_1$ represents the maximally 
amplified scale, $\eta_1=1/k_1$, and 
$g_1=k_1/a(\eta_1)/M_{\rm Planck}$ is the normalized string scale.
The axion spectral index $n_{\sigma}$ is related to the power 
which characterizes the evolution of the external dimensions by
$n_{\sigma}=2(1+\alpha)/(1-\alpha)$.
Notice that a perfect Harrison-Zel'dovich spectrum $n_{\si}=1$ requires
an isotropic evolution of the external and internal dimensions,
$\alpha=-\beta=-1/3$, or $a(t) \propto 1/b(t)$. This holds only in a 
10-dimensional space-time \cite{1}.

In the following we suppose that the contribution 
of the axion field to the background is negligible and 
that the solutions to the action (\ref{Action4})
are given by (\ref{Solu}).
Assuming that the primordial axion spectrum remains uneffected, 
at least at large scales, by
the high-energy transition to the post-big bang universe,  
the axions can be the source of the perturbations
if they play the role of ``seeds'' which, by their
gravitational field, induce fluctuations in the cosmic fluid \cite{1}.
We then suppose that the axions are first order perturbations,
interacting with the cosmic fluid only
gravitationally; the  
back-reaction of the metric perturbations on their evolution  
is second order and can be neglected.

For a universe with a given cosmic fluid, the cosmological perturbation
equations are of the form
\be
 \DD X = \SS^{\si}~, \label{diff}
\ee
where $X$ is a long vector containing all the fluid perturbation
variables, $\SS^{\si}$ is a source vector which consists of certain
combinations of the axion energy-momentum tensor, and $\DD$ is a linear
ordinary differential operator.
More concretely, we first consider a (post-big bang) 
universe consisting of cold dark matter,
baryons, photons, three types of massless neutrino, and a non zero 
cosmological constant, with a total
density parameter, $\Om=1$.

The axion field $\sigma$ is a Gaussian stochastic variable
and hence its energy-momentum tensor is quadratic in $\sigma$ and 
therefore not Gaussian; this leads to 
non Gaussian perturbations. 
Moreover, although the axion field evolves 
according to a linear equation,
its energy-momentum will not; this gives rise 
to the phenomena of decoherence\footnote{Decoherence has been 
discussed in \cite{afrg} where it has been shown 
that in the axion seeds model it 
leads to negligible deviations from the perfectly coherent approximation used here.}.

The perturbations in the dark matter and radiation components are set 
to zero in the initial conditions and are subsequently 
induced by the gravitational field of the axion. Hence,
axion seed perturbations
belong to the class of {\em isocurvature perturbations}.

Notice that in the standard (inflationary) 
{\em adiabatic} scenario the source term $\SS^{\si}$ in
Eq.~(\ref{diff}) is absent and the initial conditions for 
the perturbation variables $X$ are non vanishing linear functions of 
the perturbation in the inflaton field. This leads to completely 
linear and coherent evolution of fluctuations. 
The isocurvature nature of the perturbations in our model 
will particularly show up in 
the CMB anisotropy power spectrum giving rise to an
``isocurvature hump'' and a 
different position of the first Doppler peak for equal  
total density parameter. 

\section*{CMB anisotropies and comparison with data}

As shown in \cite{1}, CMB anisotropies seeded by axions
can lead to a flat spectrum at large scale as required by the ``old'' data.   
However, in the last two years, a peak in the CMB power spectrum at $\ell \sim 200$
as been detected by several different experiments, most recently
by BOOMERanG98 and
MAXIMA-1 \cite{debbe}.
In order for our model to be compelling, it is necessary to
compare it to these observations.  

We therefore determine the CMB anisotropies by numerically solving 
the axion field
equation in the unperturbed background geometry, 
\Eq{Evol}, during the radiation and matter dominated eras, 
by computing its energy-momentum tensor
and inserting the resulting source functions in a
Boltzmann solver. The result depends on the primordial axion 
spectral index which is the initial condition, the heritage of the 
pre-big bang phase, and which depends, on the other hand, 
on the evolution of the external and internal dimensions \cite{afrg}.

\begin{figure}[ht]
\centerline{\epsfig{file=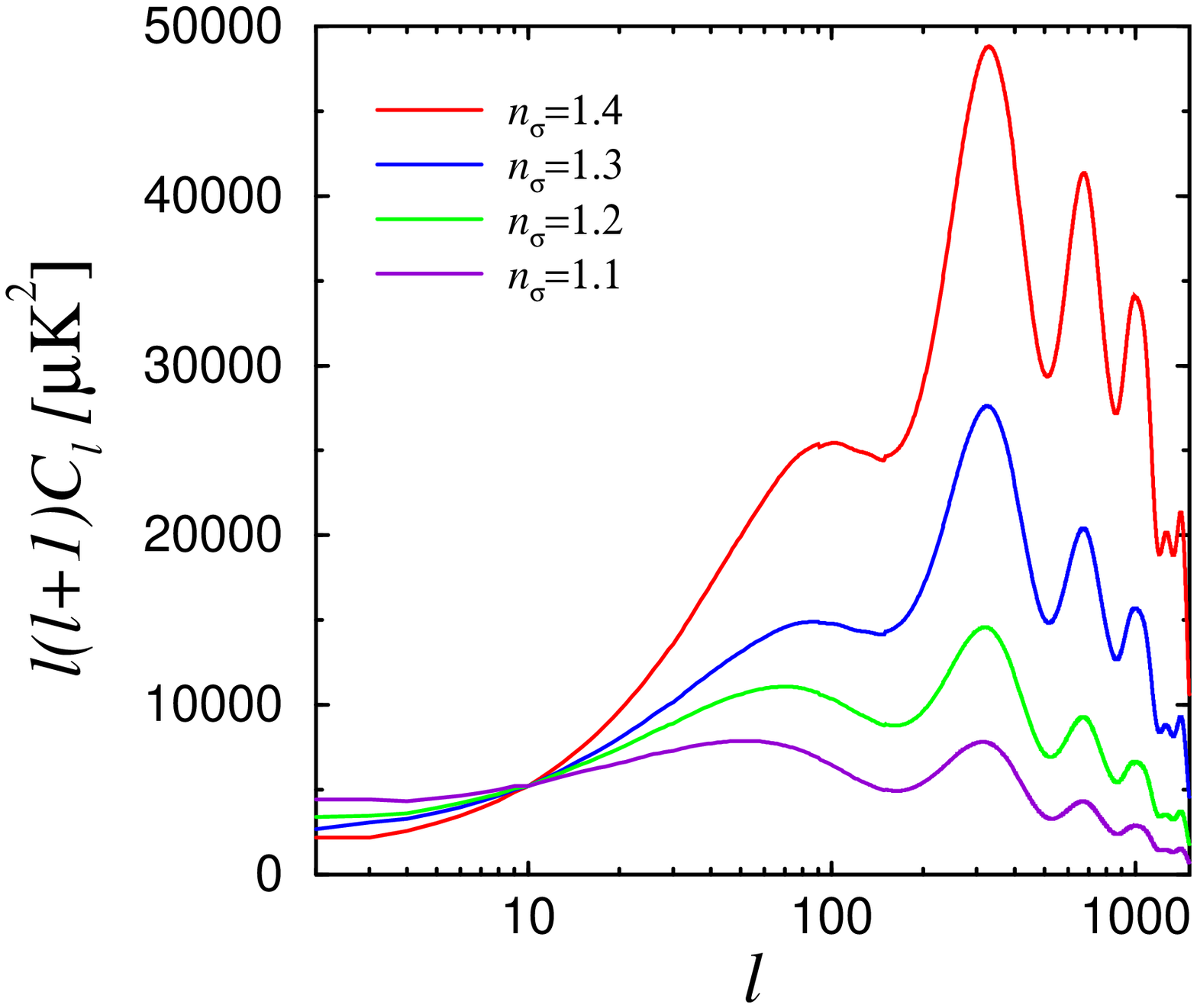, 
height=2.6in, width=3.5in } 
\epsfig{file=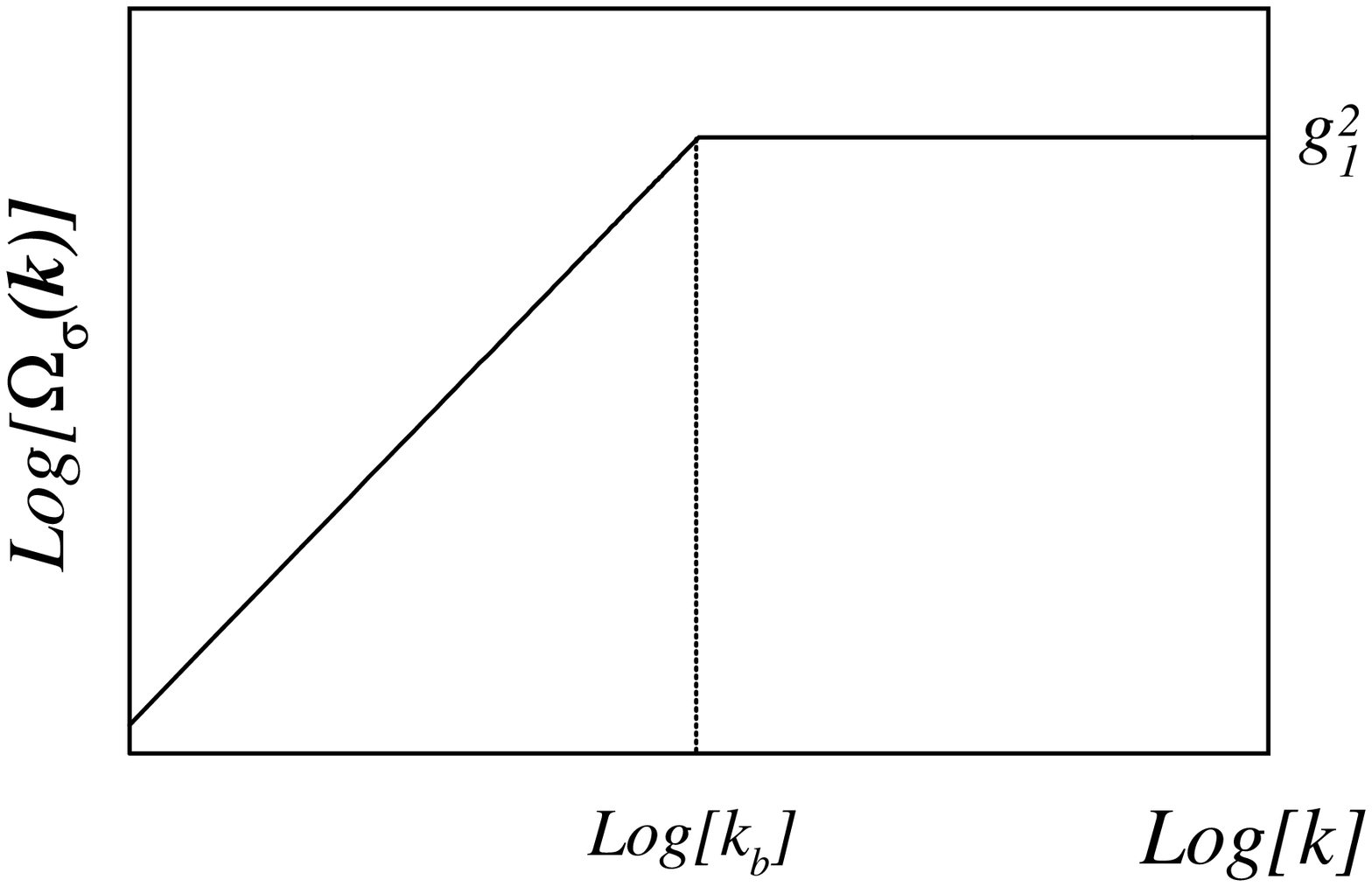, height=2.6in, width=3.1in}}
\vspace{10pt}
\caption{In the left panel, the 
CMB anisotropy power spectrum dependency on the 
primordial axion spectral index $n_{\si}$ is shown. 
We have considered a 
flat universe with $\Omega_{m}=0.3$ and $\Omega_{\Lambda}=0.7$ and 
an axionic spectral index (from top to bottom) 
$n_{\si}=1.4,~1.3,~1.2,~1.1$. In the right panel we show
the shape of the axion spectrum.}
\end{figure}

In Fig.~1 we plot the dependence of the 
CMB anisotropy power spectrum on the primordial axion spectral 
index $n_{\si}$. A slightly tilted spectral index 
($1.3 \ltapprox n_{\si} \ltapprox 1.4$) is required to have a 
sufficiently high peak.
Together with the normalization condition
of the axion spectrum imposed by Eq.~(\ref{Spettro}), namely 
$\Omega_{\si}(k_1) \simeq g_1^2$, this requires, in the simplest case, 
the presence of a kink in the spectrum. 
At very large scales the axion spectrum must be 
slightly blue,
to fit the CMB data, and only at 
an intermediate break scale, 
$k_{b}$, it must become
flat (see Fig.~1).
This requires, in terms of the evolution of the scale factors
in the pre-big bang, $\alpha \gtapprox -\beta$, 
{\em i.e.} a slower expansion of the  external 
dimensions and, correspondingly, a
somewhat faster contraction of internal dimensions at very early 
(negative) time. 

We do not want to specify the event which may have triggered such a
transition from $n_{\sigma(k < k_b)}=1+\ep$ to 
$n_{\sigma(k > k_b)}=1$ but 
it is interesting to note that
isotropic expansion and contraction 
in a $26$-dimensional space-time
gives just about the
``tilt'' needed to fit the observed CMB anisotropies
(see below).

\begin{figure}[ht]
\centerline{\epsfig{file=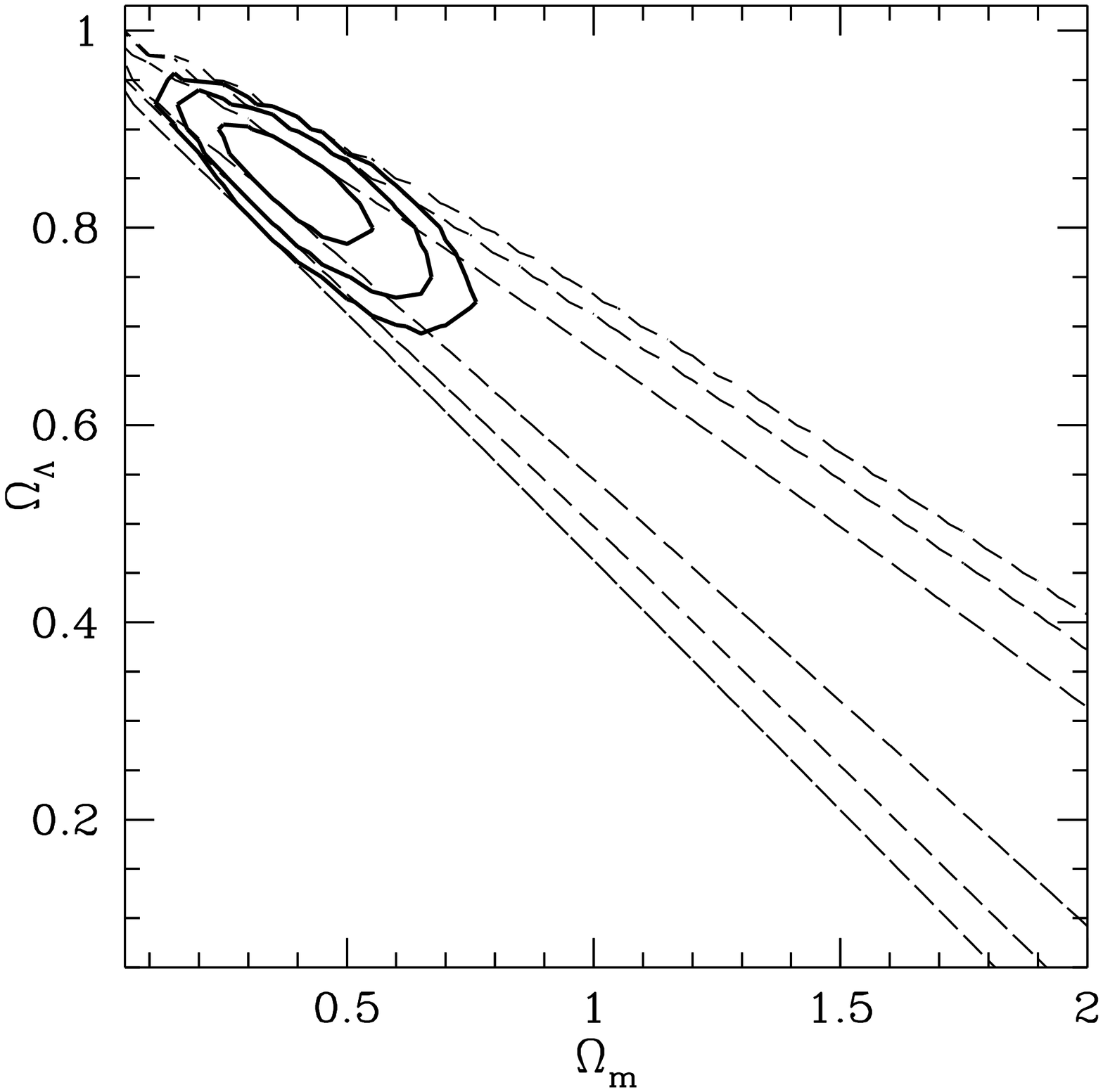, height=2.5in, width=3in} 
\epsfig{file=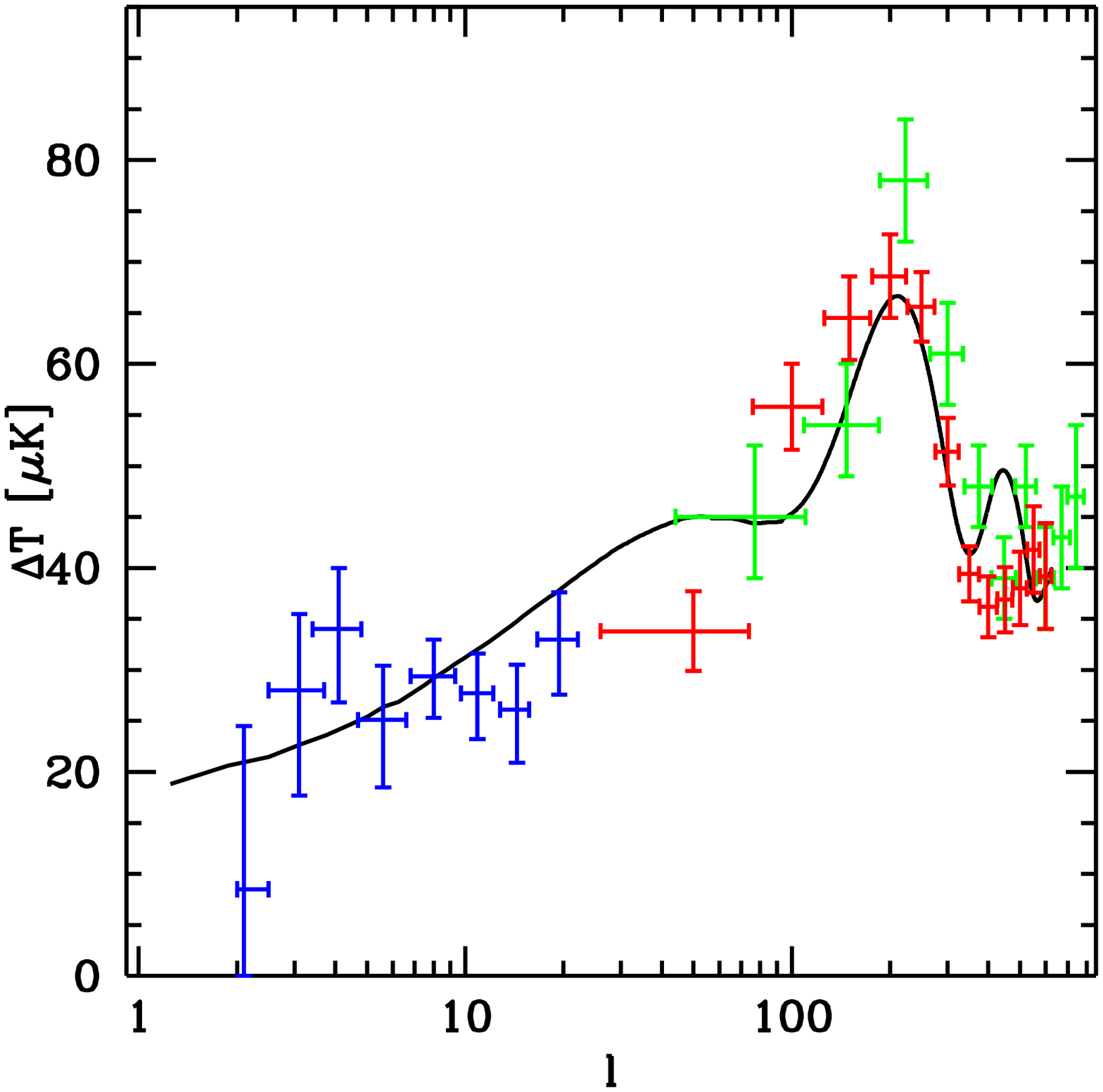, height=2.5in, width=3in}}
\vspace{10pt}
\caption{In the left panel, the $68\%$, $95\%$, 
and $99\%$ confidence levels for the 
cosmological parameters
$\Omega_{\Lambda}$ and $\Omega_{m}$, from the peak position
detected by BOOMERanG and MAXIMA-1 for the
axion model (dashed). 
The solid contours are obtained
including the SN1a data. 
In the right panel we compare
our best fit CMB anisotropy spectrum with the
COBE, MAXIMA and BOOMERanG98 data.} 
\end{figure}

As shown in Fig.~1, the CMB anisotropy power spectra 
display two characteristic isocurvature signatures which are
absent in adiabatic models. These are 
an isocurvature hump at $\ell\sim 40$ 
and the position of the first
acoustic peak at $\ell\sim 300$ for a flat universe.
While the presence of the 
hump is unavoidable, the position of the peak can be changed by 
a suitable choice of cosmological parameters.

In order to have a first peak at $\ell \sim 200$, we rescale 
our CMB anisotropy power spectrum in position, 
$\ell_{\rm rescaled}=\ell_{\rm flat}/R$, 
by  the angular diameter distance parameter for a closed universe,
$R=(1/2) \sqrt{\Omega_m/|\Omega_{K}|} \sin y$.
Here $\Om_K=1-\Om_m-\Om_\La$
is the curvature parameter and $y$ is
the following integral:
\begin{equation}
y=\sqrt{|\Omega_{K}|}\int_{0}^{z_{dec}} {dz \over 
[\Omega_{m}(1+z)^3+\Omega_{K}(1+z)^2+\Omega_{\Lambda}]^{1/2}} \label{yy}.
\end{equation}

The condition $R=$ const.\
identifies curves in the $\Omega_m - \Omega_\La$ plane,
with nearly degenerate $C_\ell$ spectra.
In  Fig.~2 we plot the confidence levels
in the $\Omega_\La - \Omega_m$ plane which are then
combined  with SN1a results \cite{SN1}. 
It is clear
that the model can be brought in reasonable agreement with 
observations only if the universe is moderately closed,
with $\Omega_{\Lambda} \sim 0.85$ and $\Omega_m \sim 0.4$,
which is also compatible with cluster abundance and X-ray data.

In  Fig.~2 we also compare the CMB anisotropy power spectrum
of our model for this choice of cosmological parameters with the COBE,
MAXIMA-1, and BOOMERanG98 data. The position of the first 
acoustic peak has been correctly adjusted; nonetheless, the width of
the peak, compressed by the increase of $R$, 
as well as the isocurvature hump, are still not in very good
agreement with the data. The 
resulting normalized $\chi^2$ is about $\sim 1.8$ for the best-fit,
which ``excludes'' the model at 70\% confidence
\footnote{One has however to keep
in mind that the $C_\ell$'s are not Gaussian and therefore the
probability for our model to lead to the measured CMB anisotropies is
even somewhat higher than 30\%.}. Clearly more and better
data  around the isocurvature
hump region, {\em i.e.} $\ell\sim 50$, is needed to decide definitely
whether the model is ruled out.

Even if our model will turn out to disagree with better data, we
believe that we learn the important lesson that cosmological
parameters obtained from  CMB anisotropies are strongly model
dependent, a point which is  swept under the carpet by the vast
majority of the circulating ``parameter-fitting'' literature. 
We believe that it is very important in the future to concentrate on model
independent quantities, like inter-peak distances, to determine
cosmological parameters.


\begin{references}
\newcommand{\plb}{{\em Phys. Lett. B}\ }
%\newcommand{\prl}{{\em Phys. Rev. Lett.}\ }
%\newcommand{\prd}{{\em Phys. Rev. D}\ }
\newcommand{\npb}{{\em Nucl. Phys. B}\ }
\newcommand{\bb}{\bibitem}

\bibitem{PBB} G.\ Veneziano,  {\em Phys. Lett. B}  {\bf 265} 287 (1991); 
	M.\ Gasperini and G.\ Veneziano,
	{\em Astropart. Phys.} {\bf 1}, 317
	(1993); {\em Mod. Phys. Lett. A} {\bf 8}, 3701 (1993); {\em
	Phys. Rev. D} {\bf 50}, 2519
	(1994).  
%An updated collection of papers on the pre-big bang
%scenario is available at {\tt http://www.to.infn.it/\~{}gasperin/}.
\bibitem{5b} M. Gasperini and M. Giovannini, \plb {\bf 282}, 36 (1991); 
	{\em Phys. Rev. D} {\bf 47}, 1519 (1993); M. Gasperini and
	G. Veneziano, Ref. \cite{1};  R. Brustein, M. Gasperini, 
	M.~Giovannini, V. F. Mukhanov and G. Veneziano, 
	{\em Phys. Rev. D} {\bf 51}, 6744 (1995).
\bibitem{Copeland} E.\ J.\ Copeland, R.\ Easther and D.\ Wands,
	{\em Phys. Rev. D} {\bf 56}, 874 (1997);
	E.\ J.\ Copeland, J.\ E.\ Lidsey and D.\ Wands,
	\npb {\bf 506}, 407 (1997). 
\bibitem{1} R. Durrer, M. Gasperini, M. Sakellariadou and G.
	Veneziano, {\em Phys. Rev. D} {\bf 59}, 043511 (1999).
\bb{debbe} P. de Bernardis et al., {\em Nature} {\bf 404}, 955 (2000); 
S. Hanany et al., {\tt astro-ph/005124}, {\em Ap. J. Lett.} submitted (2000).
\bibitem{afrg} A. Melchiorri, F. Vernizzi, R. Durrer and G. Veneziano, 
        {\em Phys. Rev. Lett.} {\bf 83}, 4464 (1999);
        F. Vernizzi, A. Melchiorri and R. Durrer, 
        {\tt astro-ph/0008232}, submitted to {\em Phys. Rev. D} (2000).
\bb{SN1} S. Perlmutter et al., {\em Nature}, {\bf 391}, 51 (1999).

\end{references}
\end{document}